\documentstyle[11pt]{article}
\def\fnote#1#2{\begingroup\def\thefootnote{#1}\footnote{#2}\addtocounter
{footnote}{-1}\endgroup}

\begin{document}

\hfill{UTTG-04-08}

\vspace{36pt}

\begin{center}
{\large {\bf {A Tree Theorem for Inflation}}}

\vspace{36pt}
Steven Weinberg\fnote{*}{Electronic address:
weinberg@physics.utexas.edu}\\
{\em Theory Group, Department of Physics, University of
Texas\\
Austin, TX, 78712}

\vspace{30pt}

\noindent
{\bf Abstract}
\end{center}
\noindent
It is shown that the generating function for tree graphs in the ``in-in'' formalism may be calculated by solving the classical equations of motion subject to certain constraints.  This theorem is illustrated by application to the evolution of a single inflaton field in a Robertson--Walker background.

 \vfill

\pagebreak

\begin{center}
{\bf I. Introduction}
\end{center}

Quantum field theory is used differently in cosmology than in elementary particle physics.  In particle physics we need to know matrix elements between ``in'' states and ``out'' states, defined respectively by their appearance at times long before and long after a collision.  In contrast, in cosmology we are interested in the expectation value of various operators in an ``in'' state, usually defined as a state that looks like the vacuum at very early times, long before perturbations leave the horizon during inflation.  To calculate such expectation values, we must use the ``in--in'' formalism of Schwinger {\em et al.}[1]  

The ``in--in'' formalism provides graphical rules for calculating expectation values, similar to but more complicated than the Feynman rules used in particle physics.  A special role in these calculations is provided by tree graphs, which in many theories give much larger contributions to expectation values than graphs with loops.  For instance, in general relativity the coupling constant $G$ appears only in a factor $1/G$ multiplying the whole Einstein--Hilbert action, so each graviton propagator yields a factor $G$ and each interaction comes with a factor $1/G$.  A connected graph with $I$ internal lines and $V$ vertices thus yields a factor $G^{I-V}=G^{L-1}$, where $L=I-V+1$ is the number of loops, so at a characteristic frequency $H$, graphs with $L$ loops are suppressed relative to tree graphs by factors of order $(GH^2)^L$.  The strength of fluctuations in the cosmic microwave background suggests that $GH^2\approx 10^{-10}$.  This suppression can partly be compensated by the appearance of powers of $\ln a$[2] 
(where $a$ is the Robertson--Walker scale factor), but for this to make loops competitive with trees the universe would have to expand after horizon exit by something like $10^{10}$ e-foldings.  For the same reason, tree graphs will dominate in any theory with a small very coupling constant $g$ that appears only as a factor $1/g$ in the action.  

In the present state of cosmology it is fortunate that tree graphs make a much larger contribution to correlation functions than graphs with loops.  During the long period from when perturbations left the horizon during inflation to when they re-entered the horizon in the radiation- or matter-dominated era, there were various events, such as reheating, lepton and baryon synthesis, and dark matter decoupling, about which we know essentially nothing.  The only reason that we are able to relate observations of the microwave background anisotropies or large scale structure to what happened in inflation is that certain quantities such as the curvature perturbation $\zeta$ and the gravitational wave amplitude are believed to be time-independent when perturbations are outside the horizon[3] --- that is, when the physical wave number $k/a$ is much less than the expansion rate $H$.  For tree graphs, when the physical wave numbers associated with external lines are all much less than $H$ then the same is true of the wave numbers associated with all internal lines but, as shown by the $\ln a$ factors mentioned above, this is not true of graphs with loops, and so the theorems that state the constancy of $\zeta$ or of the gravitational wave amplitude outside the horizon apply only to tree graphs. 

It is well known in elementary particle physics that the tree graph contributions to matrix elements between ``in'' and ``out'' states can be calculated by solving the classical field equations in the presence of an external $c$-number current.[4]  In the Appendix to a recent paper[5] I remarked that the same is true in the ``in-in'' formalism, but the prescription stated there was not correct.  Section II of the present paper states and proves a general theorem (hopefully correct this time)  for the evaluation of tree contributions to expectation values by the use of classical field equations.  In Sections III and IV this theorem is applied to inflation.  The derivation of the tree theorem in Section II relies on an analytic continuation of trajectories in the path-integral formalism, and since this is not treated rigorously, we check in Section V that this theorem does give the correct results in the first few orders of perturbation theory for the inflationary model discussed in Sections III and IV.  The methods of this paper are not really needed in the calculation of tree contributions to specific expectation values, since the rules for such calculations are already well known, but it is hoped that our results may prove useful in proving general theorems about non-Gaussian correlations in cosmology.

\begin{center}
{\bf II. Tree Theorem for the ``In-In'' Formalism}
\end{center}

We will consider a general Hamiltonian system, with Hermitian operators $Q_a(t)$ and their canonical conjugates $P_a(t)$, satisfying the usual commutation relations
\begin{equation}
[Q_a(t),P_b(t)]=i\delta_{ab}\;,~~~~[Q_a(t),Q_b(t)]=[P_a(t),P_b(t)]=0\;.
\end{equation}
In field theories $a$ is a compound index, including a spatial coordinate as well as discrete indices labeling the nature of the various operators, and $\delta_{ab}$ includes a delta function in the spatial coordinates as well as Kronecker deltas for the discrete indices.  These are Heisenberg-picture operators, with a time-dependence generated by a Hamiltonian $H[Q(t),P(t),t]$:
\begin{equation}
\dot{Q}_a(t)=-i\Big[Q_a(t),H[Q(t),P(t),t]\Big]\;,~~~~\dot{P}_a(t)=-i\Big[P_a(t),H[Q(t),P(t),t]\Big]\;.
\end{equation}
We are allowing the Hamiltonian here to have an explicit dependence on time for a reason discussed in [5]: When $Q_a(t)$ and  $P_a(t)$ are the {\em fluctuations} of canonical variables around time-dependent background values, the Hamiltonian that generates their time-dependence is not the time-independent Hamiltonian for the total canonical variables, but the sum of the terms in this Hamiltonian of second and higher order in the fluctuations, so that a time-dependence is introduced by dropping the terms of zeroth and first order in fluctuations.  The differential equations (2) show that in $H[Q(t),P(t),t]$ we can set the time argument of $Q_a(t)$ and  $P_a(t)$ equal to any fixed time $t_*$, but the explicit time dependence of $H[Q(t_*),P(t_*),t]$ still remains.

Instead of giving a formula for  expectation values of products of the $Q_a(t_1)$ at a fixed time $t_1$ in an ``in'' state $|0, {\rm in}\rangle$ defined by its appearance at a time $t_0<t_1$, we will give a formula for  a generating function $W[J,t_1]$ of a c-number current $J_a$, from which all such expectation values may be obtained.  The generating function is defined by
\begin{equation}
e^{W[J,t_1]}\equiv \left\langle 0, {\rm in}\left|\exp\Big[\sum_a Q_a(t_1)J_a\Big]\right|0, {\rm in}\right\rangle\;,
\end{equation}
and from its derivatives we can calculate the expectation value of a product of $n$ of the $Q$'s:
\begin{equation}
\left\langle 0, {\rm in}\left|Q_a(t_1)\,Q_b(t_1)\cdots\right|0, {\rm in}\right\rangle=\left[\frac{\partial^n }{\partial J_a\partial J_b\cdots}\,\exp\Big[W[J,t_1]\Big]\right]_{J=0}
\end{equation}
(In field theories, sums over $a$ like that in (3) include an integral over a spatial coordinate as well as sums over discrete indices, and the derivatives in (4) are functional derivatives.)  We need not take $J_a$ to be real, but if we do then the generating function is real also.

To calculate $W[J,t_1]$ in the tree approximation we introduce a pair of complex c-number $J$-dependent functions of time, $q_{La}(t)$ and $q_{Ra}(t)$, which are defined by three conditions:

\vspace{6pt}

\noindent
({\bf A}) Both $q_{La}(t)$ and $q_{Ra}(t)$ satisfy the Lagrangian equations of motion
\begin{equation}
\frac{d}{dt}\left(\frac{\partial L[q_L(t),\dot{q}_L(t),t]}{\partial\dot{q}_{La}(t)}\right)=\frac{\partial L[q_L(t),\dot{q}_L(t),t]}{\partial q_{La}(t)}\;,
\end{equation}
and
\begin{equation}
\frac{d}{dt}\left(\frac{\partial L[q_{R}(t),\dot{q}_{R}(t),t]}{\partial\dot{q}_{Ra}(t)}\right)=\frac{\partial L[q_{R}(t),\dot{q}_{R}(t),t]}{\partial q_{Ra}(t)} \;,
\end{equation}
where $L[q(t),\dot{q}(t),t]$ is obtained from the classical Hamiltonian $H$ by using the expression
\begin{equation}
L[q(t),\dot{q}(t),t]=\sum_a p_a(t)\dot{q}_a(t)-H[q(t),p(t),t]\;.
\end{equation}
with $p_a(t)$ eliminated from the right-hand side by  using the classical formula
\begin{equation}
\dot{q}_a(t)=\frac{\partial H[q(t),p(t),t]}{\partial p_a(t)}\;.
\end{equation}

\vspace{6pt}

\noindent
({\bf B}) The $q_{La}(t)$ and $q_{Ra}(t)$ and their time derivatives  satisfy constraints at the time $t_1$:
\begin{equation}
q_{La}(t_1)=q_{Ra}(t_1)\;,
\end{equation}
and
\begin{equation}
\frac{\partial L[q_L(t_1),\dot{q}_L(t_1),t_1]}{\partial\dot{q}_{La}(t_1)}-\frac{\partial L[q_R(t_1),\dot{q}_R(t_1),t_1]}{\partial\dot{q}_{Ra}(t_1)}  =-iJ_a\;.
\end{equation}

\vspace{6pt}

\noindent
({\bf C}) The $q_{La}(t)$ and $q_{Ra}(t)$ and their time derivatives also satisfy constraints at the time $t_0$  that is used to define the state $|0, {\rm in}\rangle$, constraints that depend on the nature of this state.  In particular, if $|0, {\rm in}\rangle$ is a state that looks like the Bunch--Davies vacuum at a time $t_0=-\infty$, then $q_{La}(t)$ and $q_{Ra}(t)$ satisfy ``positive frequency'' and ``negative frequency'' conditions, respectively; that is, they must for $t\rightarrow -\infty$   be superpositions of terms with time-dependence proportional to $e^{-i\omega t}$ and $e^{+i\omega t}$, respectively, where the $\omega$s are various positive frequencies.

With $q_{La}(t)$ and $q_{Ra}(t)$ calculated in terms of the current using these three conditions, the generating function $W[J,t_1]$ is given in the tree approximation by
\begin{equation}
W[J,t_1]_{\rm tree}=i\int_{t_0}^{t_1}dt\;\Big\{L[q_R(t),\dot{q}_R(t),t]-L[q_L(t),\dot{q}_L(t),t]\Big\}+\sum_a q_{Ra}(t_1)\,J_a\;.
\end{equation}

(There is an easy generalization of this theorem: Instead of the linear function $\sum_a Q_a(t_1)J_a$ in the exponential on the right-hand side of the definition (3) of $W$, we could insert an arbitrary function ${\cal J}[Q(t_1)]$ of all the $Q_a$ at the same time $t_1$.  Then in the tree approximation $W$ would be given by
$$
W_{\rm tree}=i\int_{t_0}^{t_1}dt\;\Big\{L[q_R(t),\dot{q}_R(t),t]-L[q_L(t),\dot{q}_L(t),t]\Big\}+{\cal J}[q_R(t_1)]\;$$
with $J_a$ on the right-hand side of the constraint (10) replaced with $\partial {\cal J}(q)/\partial q_a$, in which we set  $q_a=q_{Ra}(t_1)$.)

The proof of the tree theorem relies on the path-integral formulation of the ``in-in'' formalism, so we begin with a brief derivation of the path-integral formula for $\exp(W[J,t_1])$.  At any given time $t$, we introduce a complete set of  states $|q,t\rangle$, defined as normalized eigenstates of the $Q_a(t)$ with eigenvalues $q_a$:
\begin{equation}
Q_a(t)|q,t\rangle =q_a|q,t\rangle\;,~~~~~~\langle q,t| q',t\rangle =\prod_a \delta(q_a-q'_a)\;.
\end{equation}
The expectation value (3) may then be written
\begin{eqnarray}
e^{W[J,t_1]}&=&\int \left(\prod_a dq_a\,dq'_a\,dq''_a \right) \Psi^*_0(q')\langle q,t_1| q',t_0\rangle^*\nonumber\\&&\times  \exp \left(\sum_a q_aJ_a\right)
\langle q,t_1| q'',t_0\rangle \Psi_0(q'')\;,
\end{eqnarray}
where $\Psi_0(q)$ is the wave function of the state in which the expectation value is taken
\begin{equation}
\Psi_0(q)\equiv \langle q,t_0|0,{\rm in}\rangle\;.
\end{equation}
The matrix elements between eigenstates of the $Q$'s at  times $t_1$ and $t_0$ with $t_0<t_1$ are given by an integral over real functions $q_a(t)$ that interpolate between the eigenvalues at $t_0$ and $t_1$ together with independent unconstrained real functions $p_a(t)$: 
\begin{eqnarray}
\langle q,t_1| q',t_0\rangle&=&\int \left(\prod_{a,t} d q_a(t)\,dp_a(t)\right) \left(\prod_a\delta(q_a(t_0)- q'_a)\right)\left(\prod_a\delta(q_a(t_1)- q_a)\right)\nonumber\\&&\times\exp\left\{i\int_{t_0}^{t_1}dt\;\left[\sum_a p_a(t) \dot{q}_a(t)-H[q(t),p(t),t]\right]\right\} \;.
\end{eqnarray}
(Eq.~(15) and its derivation are the same as encountered in the familiar derivation of the path-integral formula for the S-matrix.)  Using this in Eq.~(13), we must introduce separate integration variables  $q_{La}(t)$, $p_{La}(t)$ and $q_{Ra}(t)$, $p_{Ra}(t)$ in the path-integral formulas for  $\langle q,t_1| q',t_0\rangle$ and  $\langle q,t_1| q'',t_0\rangle$, respectively.  This gives
\begin{eqnarray}
&&e^{W[J,t_1]}=\int \left(\prod_{a,t} d q_{La}(t)\,dp_{La}(t)\right)\int \left(\prod_{a,t} d q_{Ra}(t)\,dp_{Ra}(t)\right)\nonumber\\&&~~
\times \Psi_0^*\Big(q_L(t_0)\Big)\Psi_0\Big(q_R(t_0)\Big)\left(\prod_a\delta(q_{La}(t_1)- q_{Ra}(t_1))\right)
\exp \left(\sum_a q_{Ra}(t_1)J_a\right)\nonumber\\&&~~
\times \exp\left(-i\int_{t_0}^{t_1}dt\;\left\{\sum_a p_{La}(t) \dot{q}_{La}(t)-H[q_L(t),p_L(t),t]\right\}\right)\nonumber\\&&~~\times \exp\left(i\int_{t_0}^{t_1}dt\;\left\{\sum_a p_{Ra}(t) \dot{q}_{Ra}(t)-H[q_R(t),p_R(t),t]\right\}\right)\;.
\end{eqnarray}
There is no special reason why we chose $q_{Ra}(t_1)$ rather than $q_{La}(t_1)$ to multiply the current in the first exponential; the delta function makes this choice inconsequential.

Eq.~(16) leads to well-known graphical rules: Writing $H$ as the sum of a quadratic  part  and an interaction,  we expand in powers of the interaction and  $J$, and evaluate the resulting Gaussian integral as a sum of ways of pairing the $q$s and $p$s in this expansion.  Because it enters in the exponential on the left-hand side of Eq.(16),  $W[J,t_1]$ is given by a sum of {\em connected} graphs.  To isolate the connected tree graphs, we use the same trick as is used in proving a tree theorem for the S-matrix[4]: We introduce a fictitious coupling constant $g$, defining a generating function $W[J,t_1,g]$ by
\begin{eqnarray}
&&e^{W[J,t_1,g]/g}\equiv\int \left(\prod_{a,t} d q_{La}(t)\,dp_{La}(t)\right)\int \left(\prod_{a,t} d q_{Ra}(t)\,dp_{Ra}(t)\right)\nonumber\\&&~~
\times \Psi_0^*\Big(q_L(t_0)\Big)\Psi_0\Big(q_R(t_0)\Big)\left(\prod_a\delta(q_{La}(t_1)- q_{Ra}(t_1))\right)
\nonumber\\&&~~
\times \exp \left(\frac{1}{g}\sum_a q_{Ra}(t_1) J_a\right)\exp\left(-\frac{i}{g}\int_{t_0}^{t_1}dt\;\left\{\sum_a p_{La}(t) \dot{q}_{La}(t)-H[q_L(t),p_L(t),t]\right\}\right)\nonumber\\&&~~\times \exp\left(\frac{i}{g}\int_{t_0}^{t_1}dt\;\left\{\sum_a p_{Ra}(t) \dot{q}_{Ra}(t)-H[q_R(t),p_R(t),t]\right\}\right)\;.
\end{eqnarray}
By the same argument as in Section I, a connected graph with $L$ loops makes a contribution to $W[J,t_1,g]/g$ proportional to $g^{L-1}$, so $W[J,t_1,g]$ for $g\rightarrow 0$ approaches a $g$-independent limit given by the   sum of tree graphs for any value of $g$, and in particular for the physical value $g=1$:
\begin{equation}
W[J,t_1,g]\rightarrow W[J,t_1]_{\rm tree}~~~{\rm for}~~~g\rightarrow 0\;.
\end{equation}

Now, in the limit $g\rightarrow 0$, the path integrals in Eq.~(17) are dominated by complex $J$-dependent trajectories                                                                                                                                                                    of $q_{La}(t)$, $p_{La}(t)$ and $q_{Ra}(t)$, $p_{Ra}(t)$ at which the coefficient of $1/g$ in the combined exponential is stationary, so
\begin{eqnarray}
W[J,t_1]_{\rm tree}&=&\Bigg[\sum_a q_{Ra}(t_1) J_a-i\int_{t_0}^{t_1}dt\left\{\sum_a p_{La}(t) \dot{q}_{La}(t)-H[q_L(t),p_L(t),t]\right\}\nonumber\\&&+i\int_{t_0}^{t_1}dt\left\{\sum_a p_{Ra}(t) \dot{q}_{Ra}(t)-H[q_R(t),p_R(t),t]\right\}\Bigg]_{\rm staty}\;,
\end{eqnarray}
with the condition that the trajectories be stationary (indicated by the subscript ``staty'') taken subject to the constraint (9) imposed by the factor $\prod_a\delta(q_{La}(t_1)- q_{Ra}(t_1))$ in Eq.~(17), and also subject to constraints imposed by the factors $\Psi_0^*\Big(q_L(t_0)\Big)$ and $\Psi_0\Big(q_R(t_0)\Big)$, about which more later.

The functions $p_{La}(t)$ and $p_{Ra}(t)$ are unconstrained, so in Eq.~(19) we may take them to be given by the classical conditions (8) applied to both functions.  Then Eq.~(19) becomes
\begin{eqnarray}
W[J,t_1]_{\rm tree}&=&\Bigg[\sum_a q_{Ra}(t_1) J_a -i\int_{t_0}^{t_1}dt\; L[q_L(t),\dot{q}_L(t),t]\nonumber\\&&+i\int_{t_0}^{t_1}dt\; L[q_R(t), \dot{q}_R(t),t]\Bigg]_{\rm staty}\;,
\end{eqnarray}
with $L[q,\dot{q},t]$ given by using (8) to eliminate the $p$s in Eq.~(7).  
This is the same as the desired result (11), provided we can show that the trajectories for which this quantity is stationary are those described by the three above conditions (A), (B), and (C).

To implement the condition that this is stationary with respect to variations in $q_{La}(t)$ and $q_{Ra}(t)$, we note that
\begin{eqnarray}
&&\delta\Bigg[\sum_a q_{Ra}(t_1) J_a-i\int_{t_0}^{t_1}dt\; L[q_L(t),\dot{q}_L(t),t]+i\int_{t_0}^{t_1}dt\; L[q_R(t), 
\dot{q}_R(t),t]\Bigg]\nonumber\\&&=\sum_a \delta q_{Ra}(t_1) J_a\nonumber\\&&-i\int_{t_0}^{t_1}dt\sum_a\left[ \frac{\partial L[q_L(t),\dot{q}_L(t),t]}{\partial q_{La}(t)}\delta 
q_{La}(t)+\frac{\partial L[q_L(t),\dot{q}_L(t),t]}{\partial \dot{q}_{La}(t)}\delta 
\dot{q}_{La}(t)\right]\nonumber\\&&~~+i\int_{t_0}^{t_1}dt\;\sum_a\left[ \frac{\partial L[q_R(t),\dot{q}_R(t),t]}{\partial 
q_{Ra}(t)}\delta q_{Ra}(t)+\frac{\partial L[q_R(t),\dot{q}_R(t),t]}{\partial \dot{q}_{Ra}(t)}\delta 
\dot{q}_{Ra}(t)\right]\nonumber\\&&=\sum_a \delta q_{Ra}(t_1) \left(J_a+i\frac{\partial L[q_R(t_1),\dot{q}_R(t_1),t_1]}{\partial 
\dot{q}_{Ra}(t_1)}\right) -i\sum_a \delta q_{La}(t_1) \frac{\partial L[q_L(t_1),\dot{q}_L(t_1),t_1]}{\partial 
\dot{q}_{La}(t_1)}
\nonumber\\&&-i\int_{t_0}^{t_1}dt\sum_a\left[ \frac{\partial L[q_L(t),\dot{q}_L(t),t]}{\partial q_{La}(t)}-
\frac{d}{dt}\frac{\partial L[q_L(t),\dot{q}_L(t),t]}{\partial \dot{q}_a(t)}\right]\delta 
q_{La}(t)\nonumber\\&&~~+i\int_{t_0}^{t_1}dt\;\sum_a\left[ \frac{\partial L[q_R(t),\dot{q}_R(t),t]}{\partial q_{Ra}(t)}-
\frac{d}{dt}\frac{\partial L[q_R(t),\dot{q}_R(t),t]}{\partial \dot{q}_{Ra}(t)}\right]\delta q_{Ra}(t)\;.
\end{eqnarray}
The vanishing of the coefficients of $\delta q_{La}(t)$ and $\delta q_{Ra}(t)$ for $t_0<t<t_1$ yields the Lagrangian equations of motion (5) and (6).  The vanishing of the coefficients of $\delta q_{La}(t_1)$ and $\delta q_{Ra}(t_1)$, subject to the condition that $\delta q_{La}(t_1)=\delta q_{Ra}(t_1)$, yields Eq.~(10).  

Finally, we must return to the constraint on the solutions of the Lagrangian equations of motion (5) and (6) imposed by the presence in (17) of the wave functions $\Psi_0\Big(q_R(t_0)\Big)$ and $\Psi^*_0\Big(q_L(t_0)\Big)$.  As is familiar from  the path-integral calculation of the S-matrix, where $t_0\rightarrow -\infty$ and $|0, {\rm in}\rangle$ is the ``in'' vacuum  the factor $\Psi_0\Big(q_R(t_0)\Big)$ has the effect of putting a $-i\epsilon$ in the denominator of the Fourier integral for the propagator, so when Eq.~(6) is solved using this propagator as a Green's function we get ``negative frequency'' solutions $q_{Ra}(t)$ --- that is, solutions that behave for $t\rightarrow -\infty$ as a superposition of terms with time dependence $e^{i\omega t}$, where $\omega>0$.  As remarked in [5], because $L[q_L(t_1),\dot{q}_L(t_1),t_1]$ enters in the argument of the exponential in (17) with an opposite sign to $L[q_R(t_1),\dot{q}_R(t_1),t_1]$, the effect of the wave function $\Psi^*_0\Big(q_L(t_0)\Big)$ is to constrain 
$q_{La}(t)$ to have ``positive frequency'' --- that is, to behave for $t\rightarrow -\infty$ as a superposition of terms with time dependence $e^{-i\omega t}$, again with  $\omega>0$.  

This completes the proof that the functions $q_{La}(t)$ and $q_{Ra}(t)$ that should be used in Eq.~(11) to calculate the tree contribution to $W[J,t_1]$ are those satisfying the above conditions (A), (B), and (C).  In the case where $J_a$ is real, these three conditions are consistent with the result that $q_{La}(t)^*=q_{Ra}(t)$, which makes the tree approximation (11) to the generating function $W[J,t_1]$ real, as expected.

\begin{center}
{\bf III. Application to Inflation}
\end{center}

To illustrate the use of the tree theorem of Section II in cosmology, we will consider a simple semi-realistic model, in which a single scalar field evolves in an unperturbed Robertson--Walker metric $g_{\mu\nu}$ of zero spatial curvature.  We write the scalar field $\phi({\bf x},t)$ as an unperturbed term $\bar{\phi}(t)$ plus a fluctuation $\varphi({\bf x},t)$ (called $\delta\varphi({\bf x},t)$ in [5]).  The Lagrangian for the fluctuation is
\begin{eqnarray}
L&=&\int \sqrt{g}\,d^3x\;\left[-\frac{1}{2}g^{\mu\nu}\partial_\mu\varphi\partial_\nu\varphi-U(\varphi,t)\right]\nonumber\\&=&
a^3\int d^3x\;\left[\frac{1}{2}\dot{\varphi}^2-\frac{1}{2a^2}(\nabla \varphi)^2-U(\varphi,t)\right]\;,
\end{eqnarray}
where $a(t)$ is the Robertson--Walker scale factor, and in accordance with the prescription for constructing the Hamiltonian for fluctuations mentioned in Section II, the time-dependent potential $U(\varphi,t)$ for the fluctuations is a function containing only second and higher powers of $\varphi({\bf x},t)$, and given in terms of the time-independent potential $V(\phi)$ for the total scalar field by
\begin{equation}
U(\varphi,t)\equiv V\Big(\bar{\phi}(t)+\varphi\Big)-V\Big(\bar{\phi}(t)\Big)-V'\Big(\bar{\phi}(t)\Big)\varphi\;.
\end{equation}

It is straightforward to apply the results of the previous section to this theory.  Here the index $a$ is the spatial coordinate ${\bf x}$; the variable $q_a(t)$ is $\varphi({\bf x},t)$; the current $J_a$ is $J({\bf x})$; and derivatives with respect to $q_a(t)$ or $\dot{q}_a(t)$ are functional derivatives.  The Lagrangian equations (5) and (6) are here the 
Euler--Lagrange equations
\begin{equation}
\ddot{\varphi}_L+3H\dot{\varphi}_L-a^{-2}\nabla^2\varphi_L+U'(\varphi_L,t)=0\;,
\end{equation}
\begin{equation}
\ddot{\varphi}_R+3H\dot{\varphi}_R-a^{-2}\nabla^2\varphi_R+U'(\varphi_R,t)=0\;,
\end{equation}
where $H(t)\equiv \dot{a}(t)/a(t)$, the prime on $U$ indicates a derivative with respect to its field argument, and the constraints (9) and (10) read
\begin{equation}
\varphi_L({\bf x},t_1)=\varphi_R({\bf x},t_1)\;,
\end{equation}
\begin{equation}
\dot{\varphi}_L({\bf x},t_1)-\dot{\varphi}_R({\bf x},t_1)=-ia^{-3}(t_1)J({\bf x})\;.
\end{equation}
With $t_0=-\infty$ and expectation values calculated for the Bunch--Davies vacuum, the ``positive frequency'' and ``negative frequency'' constraints require in this model  that, for $t\rightarrow -\infty$,
$\varphi_L({\bf x},t)$ and $\varphi_R({\bf x},t)$ approach superpositions of $\exp(i{\bf k}\cdot{\bf x}-ik\eta)$ and 
$\exp(i{\bf k}\cdot{\bf x}+ik\eta)$, respectively, where $\eta$ is the conformal time, with $\dot{\eta}>0$.

After solving Eqs.~(24) and (25) subject to these constraints, the generating functional $W[J,t_1]$ can be calculated in the tree approximation from Eq.~(11), which for this model reads
\begin{equation}
W[J,t_1]_{\rm tree}=i\int_{-\infty}^{t_1}dt\Big\{L[\varphi_R(t),\dot{\varphi}_R(t),t]-L[\varphi_L(t),\dot{\varphi}_L(t),t]\Big\}+\int d^3x\;\varphi_R({\bf x},t_1)\,J({\bf x})\,
\end{equation}
with the Lagrangian $L$ given by Eq.~(22).  The integral of the Lagrangian can be simplified by integrating by parts and then using the field equation.  Note that
\begin{eqnarray*}
&&\int_{-\infty}^{t_1}dt\int d^3x\; a^3(t)\left[\frac{1}{2}\dot{\varphi}^2-\frac{1}{2a^2}(\nabla \varphi)^2-U(\varphi,t)\right]
=\frac{1}{2}a^3(t_1)\varphi(t_1)\dot{\varphi}(t_1)\nonumber\\&&~~-
\int_{-\infty}^{t_1}dt\int d^3x\; a^3(t)\left[\frac{1}{2}\varphi\Big(\ddot{\varphi}+3H\dot{\varphi}-\frac{1}{a^2}\nabla^2 \varphi\Big)+U(\varphi,t)\right]\;.
\end{eqnarray*}
(There is no contribution from the lower limit of the integral, because the integrand oscillates increasingly rapidly for $t\rightarrow -\infty$.)  Hence by using the field equations (24) and (25), Eq.~(28) becomes
\begin{eqnarray*}
&& W[J,t_1]_{\rm tree}=\frac{i}{2}a^3(t_1)\int d^3x\;\left\{\varphi_R({\bf x}, t_1)\dot{\varphi}_R({\bf x}, t_1)-\varphi_L({\bf x}, t_1)\dot{\varphi}_L({\bf x}, t_1)\right\}\nonumber\\&&+i\int_{-\infty}^{t_1}dt\;a^3(t)\,\int  d^3x\;\Bigg\{\frac{1}{2}\varphi_R({\bf x}, t)\,U'\Big(\varphi_R({\bf x}, t),t\Big)-U\Big(\varphi_R({\bf x}, t),t\Big)\nonumber\\&&~~~~ - \frac{1}{2}\varphi_L({\bf x}, t)\,U'\Big(\varphi_L({\bf x}, t),t\Big)+U\Big(\varphi_L({\bf x}, t),t\Big)\Bigg\}
\nonumber\\&&
+\int d^3x\;\varphi_R({\bf x},t_1)\,J({\bf x})\;.
\end{eqnarray*}
Using the constraints (26) and (27), we see that the first  term is $-1/2$ the last term, so 
\begin{eqnarray}
&& W[J,t_1]_{\rm tree}=i\int_{-\infty}^{t_1}dt\;a^3(t)\,\int d^3x\;\Bigg\{\frac{1}{2}\varphi_R({\bf x}, t)U'\Big(\varphi_R({\bf x}, t),t\Big)-U\Big(\varphi_R({\bf x},t),t\Big)\nonumber\\&&~~~~ - \frac{1}{2}\varphi_L({\bf x}, t)U'\Big(\varphi_L({\bf x},t),t\Big)+U\Big(\varphi_L({\bf x},t),t\Big)\Bigg\}
\nonumber\\&&
+\frac{1}{2}\int d^3x\;\varphi_R({\bf x},t_1)\,J({\bf x})\;.
\end{eqnarray}
This is the form we will use in what follows.  With $W$ calculated in this way, the tree approximation to the expectation value of a product of $n$ $\varphi$s is given by Eq.~(4) as
\begin{equation}
\left\langle 0, {\rm in}\left|\varphi({\bf x},t_1)\,\varphi({\bf y},t_1)\cdots\right|0, {\rm in}\right\rangle_{\rm tree}=\left[\frac{\delta^n }{\delta J({\bf x})\delta J({\bf y})\cdots}\,\exp\Big[W[J,t_1]_{\rm tree}\Big]\right]_{J=0}
\end{equation}

\begin{center}
{\bf IV. Integral Equation Formulation}
\end{center}

We will now re-write the Euler--Lagrange equations as integral equations that incorporate  the constraints on the behavior of $\varphi_L({\bf x},t)$ and $\varphi_R({\bf x},t)$ for $t\rightarrow - \infty$ as well as the constraints (26) and (27).
It is these integral equations that will be used in the following section to check that this formalism generates the usual results of perturbation theory.

We first separate the potential $U$ into a term quadratic in $\varphi$ and an interaction term $\Gamma$ containing only cubic and higher terms:
\begin{equation}
U(\varphi,t)\equiv \frac{1}{2}\varphi^2\,U''(0,t)+\Gamma(\varphi,t)\;.
\end{equation}
We introduce functions $u_k(t)$ for which  $u_k(t)\exp(i{\bf k}\cdot{\bf x})$ are ``positive frequency'' solutions of the linearized Euler-Lagrange equations, --- that is,
\begin{equation}
\ddot{u}_k(t)+3H(t)\dot{u}_k(t)+\Big(k/a(t)\Big)^2 u_k(t)+U''(0,t)u_k(t)=0\;.
\end{equation}
with ``positive frequency'' interpreted to mean that  the WKB solution for $t\rightarrow -\infty$ is proportional to $$a(t)^{-1}\exp\left(-ik\int^t dt'/a(t')\right)\;,$$ rather than to its complex conjugate.  It will be convenient to normalize these functions so that for $t\rightarrow -\infty$ the WKB solution takes the form
\begin{equation}
u_k(t)\rightarrow \frac{1}{(2\pi)^{3/2}\sqrt{2k}\,a(t)}\exp\left(-ik\int^t_{t_*}\frac{dt'}{a(t')}\right)\;,
\end{equation}
with $t_*$ any fixed time.  The Wronskian of $u_k$ and $u_k^*$ is proportional for all times to $1/a^3$, so  the normalization (33)  gives it the value
\begin{equation}
u_k(t)\dot{u}_k^*(t)-\dot{u}_k(t) u_k^*(t)=\frac{i}{(2\pi)^3 a^3(t)}\;.
\end{equation}
Using these functions, we can construct a Green's function:
\begin{equation}
G({\bf x}-{\bf x}',t,t')\equiv \theta(t-t')\Big(G_0({\bf x}-{\bf x}',t,t')+G_0^*({\bf x}-{\bf x}',t,t')\Big)\;.
\end{equation}
where $\theta$ is the usual step function, and 
\begin{equation}
G_0({\bf x}-{\bf x}',t,t')\equiv -i\int d^3k \;\exp\Big(i{\bf k}\cdot({\bf x}-{\bf x}')\Big)\,u_k(t)u_k^*(t')\;.
\end{equation}
Both $G_0$ and its complex conjugate satisfy the homogeneous wave equation
\begin{equation}
\left(\frac{\partial^2}{\partial t^2}+3H(t)\frac{\partial}{\partial t}-a^{-2}(t)\nabla^2 +U''(0,t)\right)G_0({\bf x}-{\bf x}',t,t')=0
\end{equation}
and the Wronskian formula (34) then tells us that $G$ satisfies the corresponding inhomogeneous equation:
\begin{equation}
\left(\frac{\partial^2}{\partial t^2}+3H(t)\frac{\partial}{\partial t}-a^{-2}(t)\nabla^2 +U''(0,t)\right)G({\bf x}-{\bf x}',t,t')=-\delta(t-t')\delta^3({\bf x}-{\bf x}')\;.
\end{equation}
Solutions of Eqs.~(24) and (25) that  satisfy the constraints (26) and (27) and the positive and negative frequency conditions, respectively, are given by the coupled integral equations
\begin{eqnarray}
&&\varphi_L({\bf x},t)=\int d^3x'\int_{-\infty}^{t_1} dt'\;\Bigg\{G({\bf x}-{\bf x}',t,t')\,a^3(t')\,\Gamma'\Big(\varphi_L({\bf x}',t'),t'\Big)\nonumber\\&&~~~+G_0({\bf x}-{\bf x}',t,t')\,a^3(t')\,\Bigg[\Gamma'\Big(\varphi_R({\bf x}',t'),t'\Big)-\Gamma'\Big(\varphi_L({\bf x}',t'),t'\Big)\Bigg]\Bigg\}\nonumber\\&&~~~+i\int d^3x'\;G_0({\bf x}-{\bf x}',t,t_1)\,J({\bf x}')\;,
\end{eqnarray}
\begin{eqnarray}
&&\varphi_R({\bf x},t)=\int d^3x'\int_{-\infty}^{t_1} dt'\;\Bigg\{G({\bf x}-{\bf x}',t,t')\,a^3(t')\,\Gamma'\Big(\varphi_R({\bf x}',t'),t'\Big)\nonumber\\&&~~~-G^*_0({\bf x}-{\bf x}',t,t')\,a^3(t')\,\Bigg[\Gamma'\Big(\varphi_R({\bf x}',t'),t'\Big)-\Gamma'\Big(\varphi_L({\bf x}',t'),t'\Big)\Bigg]\Bigg\}\nonumber\\&&~~~-i\int d^3x'\;G^*_0({\bf x}-{\bf x}',t,t_1)\,J({\bf x}')\;.
\end{eqnarray}

Functions $\varphi_L({\bf x},t)$ and $\varphi_R({\bf x},t)$ that satisfy these integral equations will satisfy the field equations (24) and (25) because of the properties (37) and (38) of the Green's functions, and they will satisfy the positive and negative frequency conditions because $G_0$ and $G_0^*$ respectively satisfy these conditions, while $G({\bf x}-{\bf x}',t,t')\rightarrow 0$ for $t\rightarrow -\infty$.  To check the constraint (26), and for future reference, we note that since $\theta(t_1-t')=1$, 
\begin{eqnarray}
&&\varphi_L({\bf x},t_1)=\varphi_R({\bf x},t_1)\nonumber\\&&=\int d^3x'\int_{-\infty}^{t_1}dt'\Bigg\{G_0({\bf x}-{\bf x}',t_1,t')\,a^3(t')\,\Gamma'\Big(\varphi_R({\bf x}',t'),t'\Big)\nonumber\\&&~~~~+G^*_0({\bf x}-{\bf x}',t_1,t')\,a^3(t')\,\Gamma'\Big(\varphi_L({\bf x}',t'),t'\Big)\Bigg\}\nonumber\\&&
+\int d^3x' D({\bf x}-{\bf x}',t_1)\,J({\bf x}')\;,
\end{eqnarray}
where $D$ is the real function
\begin{eqnarray}
&&D({\bf x}-{\bf x}',t_1)=iG_0({\bf x}-{\bf x}',t_1,t_1)=-iG^*_0({\bf x}-{\bf x}',t_1,t_1)\nonumber\\&&
~~=\int d^3k\;\left|u_k(t_1)\right|^2\,\exp\Big(i{\bf k}\cdot({\bf x}-{\bf x}')\Big)\;.
\end{eqnarray}
To check the constraint (27), we note that $G_0({\bf x}-{\bf x}',t,t)$ is imaginary, so 
\begin{equation}
\dot{G}({\bf x}-{\bf x}',t,t')=\theta(t-t')\,\Big(\dot{G}_0({\bf x}-{\bf x}',t,t')+\dot{G}_0^*({\bf x}-{\bf x}',t,t')\Big)\;,
\end{equation}
the dot indicating a derivative with respect to the {\em first} time argument.
It follows then from Eqs. (39) and (40) that
\begin{equation}
\dot{\varphi}_L({\bf x},t_1)-\dot{\varphi}_R({\bf x},t_1)=i\int d^3x'\;\Big(\dot{G}_0({\bf x}-{\bf x}',t_1,t_1)+\dot{G}^*_0({\bf x}-{\bf x}',t_1,t_1)\Big)\;J({\bf x}')\;.
\end{equation}
Using the Wronskian formula (34) again gives 
\begin{equation}
\dot{G}_0({\bf x}-{\bf x}',t_1,t_1)+\dot{G}^*_0({\bf x}-{\bf x}',t_1,t_1)
=-a^{-3}(t_1)\,\delta^3({\bf x}-{\bf x}')\;,
\end{equation}
so Eq. (44) shows that the constraint (27) also is satisfied.

For a real current $J({\bf x})$, and a potential $U(\varphi)$ that satisfies the reality condition $U^*(\varphi)=U(\varphi^*)$, the iterative solution of Eqs.~(39) and (40) obviously satisfies the condition $\varphi_L({\bf x},t)^*=\varphi_R({\bf x},t)$.  For $t=t_1$, when $\varphi_L({\bf x},t_1)=\varphi_R({\bf x},t_1)$, this implies that $\varphi_R({\bf x},t_1)$ is real.  It then follows that the formula (29) gives $W[J,t_1]$ real in the tree approximation.

\begin{center}
{\bf V.  Perturbation Theory}
\end{center}

The derivation  of the tree theorem in Section II was something less than mathematically rigorous.  In particular, in evaluating the path integral in the limit $g\rightarrow 0$ by setting the argument of the exponential equal to its value where stationary, we distorted the contour of integration of the variables $q_a(t)$ and $p_a(t)$ away from the real axis, without proving the validity of this analytic continuation.   As a check on the validity of this theorem, in this section we  will verify that when we expand the integral equations derived in the previous section in powers of the interaction $\Gamma$, we obtain the first few orders of perturbation theory given by the usual path-integral formulation of the ``in-in'' formalism.  

Let us first recall the diagrammatic rules given by perturbation theory for the expectation value
\begin{equation}
 \langle 0,{\rm in}|\varphi({\bf x},t_1)\varphi({\bf y},t_1)\cdots|0,{\rm in}\rangle \;,
\end{equation}
(as derived for instance in [5]), in the special case of the model discussed in Sections III and IV.  We sum contributions from diagrams containing any number of vertices, each of which can be either of $L$ or $R$ type, and is labeled with a spatial coordinate and a time $t\leq t_1$.  For each vertex of $L$ or $R$ type we include a factor $+ia^3(t)$ or $-ia^3(t)$, respectively, together with a time-dependent coupling constant  equal to the $n$-th derivative with respect to $\varphi$ of $\Gamma(\varphi,t)$ at $\varphi=0$ for a vertex to which is attached $n$ internal and/or external lines.    Attached to the vertices are external lines, corresponding to  fields in the expectation value (46), and/or internal lines connecting pairs of  vertices.  The contribution of an external line corresponding to a field $\varphi({\bf x},t_1)$ attached to a $L$ or $R$ vertex labeled with spacetime coordinates ${\bf y}$, $t$, is  a function 
\begin{equation}
L:~~~\langle 0|\varphi^I({\bf y},t)\varphi^I({\bf x},t_1)|0\rangle
\end{equation}
or
\begin{equation}
R:~~~~ \langle 0|\varphi^I({\bf x},t_1)\varphi^I({\bf y},t)|0\rangle\;,
\end{equation}
 respectively, where 
$\varphi^I({\bf x},t)$ is the interaction picture field
\begin{equation}
\varphi^I({\bf x},t)\equiv \int d^3 k\;\left[e^{i{\bf k}\cdot{\bf x}}u_k(t)\alpha({\bf k})+
e^{-i{\bf k}\cdot{\bf x}}u^*_k(t)\alpha^*({\bf k})\right]
\end{equation}
with $\alpha({\bf k})$ and $\alpha^*({\bf k})$ the usual annihilation and creation operators, and $|0\rangle$ is the bare vacuum, annihilated by $\alpha({\bf k})$.  The contribution of an internal line connecting two vertices of $L$ and/or $R$ type labeled with spacetime coordinates ${\bf x}$, $t$ and ${\bf x}'$, $t'$ is
\begin{eqnarray}
&LL:~~~~&\langle 0|\bar{T}\{\varphi^I({\bf x},t)\,\varphi^I({\bf x}',t')\}|0\rangle\;,
\\
&RR:~~~~&\langle 0|T\{\varphi^I({\bf x},t)\,\varphi^I({\bf x}',t')\}|0\rangle\;,
\\
&LR:~~~~&\langle 0|\varphi^I({\bf x},t)\,\varphi^I({\bf x}',t')|0\rangle\;,
\\
&RL:~~~~&\langle 0|\varphi^I({\bf x}',t')\,\varphi^I({\bf x},t)|0\rangle\;,
\end{eqnarray}
where $T$ and $\bar{T}$ denote time-ordered and anti-time-ordered products, respectively.
In addition, there may be lines passing through the diagram without interaction, corresponding to pairings of fields in the product whose expectation value is being calculated.  Each such line, corresponding to fields $\varphi({\bf x},t_1)$ and $\varphi({\bf y},t_1)$, makes a contribution
\begin{equation}
\langle 0|\varphi^I({\bf x},t_1)\,\varphi^I({\bf y},t_1)|0\rangle\;,
\end{equation}
the order of the fields here being unimportant, since fields at the same time commute.  Finally, we are to integrate the spatial coordinates associated with each vertex over all space, and integrate the corresponding time coordinates from $-\infty$ to $t_1$.

As a first step in checking the agreement of our theorem with these graphical rules,  we note that the quadratic term in $U(\varphi)$ drops out in the combination $\frac{1}{2}\varphi U'(\varphi,t)-U(\varphi,t)$, so Eq.~(29) may be written in terms of the interaction $\Gamma$ defined by Eq.~(31):
\begin{eqnarray}
&& W[J,t_1]_{\rm tree}=i\int_{-\infty}^{t_1}dt\int d^3x\;\Bigg\{\frac{1}{2}\varphi_R({\bf x}, t)\Gamma'\Big(\varphi_R({\bf x}, t),t\Big)-\Gamma\Big(\varphi_R({\bf x},t),t\Big)\nonumber\\&&~~~~ - \frac{1}{2}\varphi_L({\bf x}, t)\Gamma'\Big(\varphi_L({\bf x},t),t\Big)+\Gamma\Big(\varphi_L({\bf x},t),t\Big)\Bigg\}\,a^3(t)
\nonumber\\&&
+\frac{1}{2}\int d^3x\;\varphi_R({\bf x},t_1)\,J({\bf x})\;.
\end{eqnarray}
We can now expand in powers of $\Gamma$.

\vspace{6pt}

\noindent
{\bf Zeroth Order}

The zeroth order approximation to $\varphi_R({\bf x},t_1)$ is given by dropping the terms involving $\Gamma'$ in Eq.~(41):
\begin{equation}
\varphi^{(0)}_R({\bf x},t_1)=\int d^3x\;D({\bf x}-{\bf x}',t_1)\,J({\bf x}')\;.
\end{equation}
Using this in Eq.~(55) gives the zeroth order perturbation to the generating function:
\begin{equation}
W^{(0)}[J,t_1]_{\rm tree}=\frac{1}{2}\int d^3x\int d^3x'\;D({\bf x}-{\bf x}',t_1)\,J({\bf x})\,J({\bf x}')\;.
\end{equation}
The result that this is quadratic in $J$ corresponds to the elementary observation that in the absence of interactions all fluctuations are Gaussian.  The two-point correlation is given in the tree approximation by Eqs.~(30) and (57) as
\begin{equation}
\langle 0,{\rm in}|\varphi({\bf x},t_1)\,\varphi({\bf y},t_1)|0,{\rm in}\rangle_{\rm tree}=D({\bf x}-{\bf y},t_1)\;.
\end{equation}
This agrees with the above perturbative rules, because the two-point function (54) is just $ D({\bf x}-{\bf y},t_1)$.

\vspace{6pt}

\noindent
{\bf First Order}

To first order in $\Gamma$, Eq.~(55) gives the generating function as
\begin{eqnarray}
&& W^{(1)}[J,t_1]_{\rm tree}=i\int_{-\infty}^{t_1}dt\int d^3x\;\Bigg\{\frac{1}{2}\varphi^{(0)}_R({\bf x}, t)\Gamma'\Big(\varphi^{(0)}_R
({\bf x}, t),t\Big)-\Gamma\Big(\varphi^{(0)}_R({\bf x},t),t\Big)\nonumber\\&&~~~~ - \frac{1}{2}\varphi^{(0)}_L({\bf x}, t)\Gamma'\Big(\varphi^{(0)}_L({\bf x},t),t\Big)+\Gamma\Big(\varphi^{(0)}_L({\bf x},t),t\Big)\Bigg\}\,a^3(t)
\nonumber\\&&
~~~~~+\frac{1}{2}\int d^3x\;\varphi^{(1)}_R({\bf x},t_1)\,J({\bf x})\,
\end{eqnarray}
with superscripts $(0)$ and $(1)$ indicating terms of zeroth and first order in $\Gamma$.  Eqs.~(39) and (40) give
the zeroth order fields for general time as
\begin{equation}
\varphi_L^{(0)}({\bf x},t)=i\int d^3x'\;G_0({\bf x}-{\bf x}',t,t_1)\,J({\bf x}')\;,
\end{equation}
\begin{equation}
\varphi_R^{(0)}({\bf x},t)=-i\int d^3x'\;G^*_0({\bf x}-{\bf x}',t,t_1)\,J({\bf x}')\;,
\end{equation}
while from Eq.~(41) we find the first-order fields at time $t_1$:
\begin{eqnarray}
&&\varphi^{(1)}_L({\bf x},t_1)=\varphi^{(1)}_R({\bf x},t_1)\nonumber\\&&=\int d^3x'\int_{-\infty}^{t_1}dt'\;\Bigg\{G_0({\bf x}-{\bf x}',t_1,t')\,a^3(t')\,\Gamma'\Big(\varphi^{(0)}_R({\bf x}',t'),t'\Big)\nonumber\\&&~~~~+G^*_0({\bf x}-{\bf x}',t_1,t')\,a^3(t')\,\Gamma'\Big(\varphi^{(0)}_L({\bf x}',t'),t'\Big)\Bigg\}\;.
\end{eqnarray}
From the definition (36), we see that $G_0({\bf x}-{\bf x}',t,t')=-G^*_0({\bf x}-{\bf x}',t',t)$, so Eqs.~(60)--(62) give the final term in Eq.~(59) as
\begin{eqnarray*}
&&\frac{1}{2}\int d^3x\;\varphi^{(1)}_R({\bf x},t_1)\,J({\bf x})=-\frac{1}{2}\int d^3x\int d^3x'\int_{-\infty}^{t_1}dt'\;J({\bf x})\nonumber\\&&~~~~\times\Bigg\{G^*_0({\bf x}-{\bf x}',t',t_1)\,a^3(t')\,\Gamma'\Big(\varphi^{(0)}_R({\bf x}',t'),t'\Big)\nonumber\\&&~~~~+G_0({\bf x}-{\bf x}',t',t_1)\,a^3(t')\,\Gamma'\Big(\varphi^{(0)}_L({\bf x}',t'),t'\Big)\Bigg\}\nonumber\\&&=
-\frac{i}{2}\int d^3x' \int_{-\infty}^{t_1}dt'\Bigg\{\varphi_R^{(0)}({\bf x}',t_1)\Gamma'\Big(\varphi^{(0)}_R({\bf x}',t'),t'\Big)\nonumber\\&&~~~~~~~~~~~
-\varphi_L^{(0)}({\bf x}',t_1)\Gamma'\Big(\varphi^{(0)}_L({\bf x}',t'),t'\Big)\Bigg\}\;.
\end{eqnarray*}
This cancels the $\Gamma'$  terms in the first two lines of Eq.~(59), leaving us with the simple first-order result
\begin{equation}
 W^{(1)}[J,t_1]_{\rm tree}=i\int_{-\infty}^{t_1}dt\int d^3x\;\Bigg\{-\Gamma\Big(\varphi^{(0)}_R({\bf x},t),t\Big) +\Gamma\Big(\varphi^{(0)}_L({\bf x},t),t\Big)\Bigg\}\,a^3(t)\;.
\end{equation}
  According to Eqs.~(60) and (61), when we take the functional derivative of  $\varphi^{(0)}_L({\bf y},t)$ or $\varphi^{(0)}_R({\bf y},t)$ with respect to $J({\bf x})$ we get a factor $iG_0({\bf y}-{\bf x},t,t_1)$ or $-iG^*_0({\bf y}-{\bf x},t,t_1)$, respectively, while the definitions (36) and (49) give 
\begin{equation}
 iG_0({\bf y}-{\bf x},t,t_1)=\int d^3k \;\exp\Big(i{\bf k}\cdot({\bf y}-{\bf x})\Big)\,u_k(t)u_k^*(t_1)=\langle 0|\varphi^I({\bf y},t) \varphi^I({\bf x},t_1)|0\rangle \;,
\end{equation}
\begin{equation}
 -iG^*_0({\bf y}-{\bf x},t,t_1)=\int d^3k\; \exp\Big(i{\bf k}\cdot({\bf x}-{\bf y})\Big)\,u_k(t_1)u_k^*(t)=\langle 0|\varphi^I({\bf x},t_1) \varphi^I({\bf y},t)|0\rangle\;,
\end{equation}
which are the same as the factors (47) and (48) associated according to the usual graphical rules with external lines.
Thus, Eq.~(63) is just what is given by the ``in-in'' formalism for a  diagram with a single vertex, which can be $L$ or $R$ type, and any number $n\geq 3$ of external lines for which there is a term in $\Gamma$ with $n$ field factors.  For instance, in the case $n=3$, the third derivative of Eq.~(63) gives 
\begin{eqnarray*}
&&\langle 0,{\rm in}|\varphi({\bf x},t_1)\,\varphi({\bf x}',t_1)\,\varphi({\bf x}'',t_1)|0,{\rm in} \rangle = i\int_{-\infty}^{t_1}dt\int d^3y\;\Gamma'''(0,t)\,a^3(t)
\nonumber\\&& 
\times\Bigg\{G_0({\bf y}-{\bf x},t,t_1)G_0({\bf y}-{\bf x}',t,t_1)G_0({\bf y}-{\bf x}'',t,t_1)\nonumber\\&&~~~+
G^*_0({\bf y}-{\bf x},t,t_1)G^*_0({\bf y}-{\bf x}',t,t_1)G^*_0({\bf y}-{\bf x}'',t,t_1)\Bigg\}\;,
\end{eqnarray*}
which is the same as given by a graph with a single vertex to which are attached three external lines.
This three-point function has been calculated in a different way, by a perturbative solution for the Heisenberg picture fields,  by 
Seery, Malik, and Lyth[6].  It  is the analog of the three-point function calculated in a more realistic model by Maldacena[3].

\vspace{6pt}

\noindent
{\bf Second Order}

So far, we have confirmed in first order that we get the right contributions from external lines, but to check that we get the right propagators for internal lines, we have to go to second order in $\Gamma$.  According to Eq.~(55), the contribution to $W[J,t_1]_{\rm tree}$ of second order in $\Gamma$ is
\begin{eqnarray}
&& W^{(2)}[J,t_1]_{\rm tree}=i\int_{-\infty}^{t_1}dt\,a^3(t)\int d^3x\;\Bigg\{-\frac{1}{2}\varphi^{(1)}_R({\bf x}, t)\Gamma'\Big(\varphi^{(0)}_R({\bf x}, t),t\Big)\nonumber\\&&~~~~~~~~~~+\frac{1}{2}\varphi^{(0)}_R({\bf x}, t)\varphi^{(1)}_R({\bf x}, t)\Gamma''\Big(\varphi^{(0)}_R({\bf x},t),t\Big)\nonumber\\&&~~~~ +\frac{1}{2}\varphi^{(1)}_L({\bf x}, t)\Gamma'\Big(\varphi^{(0)}_L({\bf x}, t),t\Big)-\frac{1}{2}\varphi^{(0)}_L({\bf x}, t)\varphi^{(1)}_L({\bf x}, t)\Gamma''\Big(\varphi^{(0)}_L({\bf x},t),t\Big) \Bigg\}
\nonumber\\&&
~~~+\frac{1}{2}\int d^3x\;\varphi^{(2)}_R({\bf x},t_1)\,J({\bf x})\;.
\end{eqnarray}
The second-order contribution to $\varphi_R({\bf x},t_1)$ is given by Eq.~(41) as
\begin{eqnarray}
&&\varphi^{(2)}_R({\bf x},t_1)=\int d^3x'\int_{-\infty}^{t_1}dt'\Bigg\{G_0({\bf x}-{\bf x}',t_1,t')\,a^3(t')\,\Gamma''\Big(\varphi^{(0)}_R({\bf x}',t'),t'\Big)\,\varphi^{(1)}_R({\bf x}',t')\nonumber\\&&~~~~+G^*_0({\bf x}-{\bf x}',t_1,t')\,a^3(t')\,\Gamma''\Big(\varphi^{(0)}_L({\bf x}',t'),t'\Big)\,\varphi^{(1)}_L({\bf x}',t')\Bigg\}\;.
\end{eqnarray}
By again using the relation $G_0({\bf x}-{\bf x}',t,t')=-G^*_0({\bf x}-{\bf x}',t',t)$ and Eqs.~(60)--(62), now together with Eq.~(67), we see that the final term in Eq.~(66) is
\begin{eqnarray*}
&&\frac{1}{2}\int d^3x\;\varphi^{(2)}_R({\bf x},t_1)\,J({\bf x})=-\frac{1}{2}\int d^3x\int d^3x'\int_{-\infty}^{t_1}dt'\;J({\bf x})\nonumber\\&&~~~~\times\Bigg\{G^*_0({\bf x}-{\bf x}',t',t_1)\,a^3(t')\,\Gamma''\Big(\varphi^{(0)}_R({\bf x}',t'),t'\Big)\varphi^{(1)}_R({\bf x}',t')\nonumber\\&&~~~~+G_0({\bf x}-{\bf x}',t',t_1)\,a^3(t')\,\Gamma''\Big(\varphi^{(0)}_L({\bf x}',t'),t'\Big)\varphi^{(1)}_R({\bf x}',t')\Bigg\}\nonumber\\&&=
-\frac{i}{2}\int d^3x' \int_{-\infty}^{t_1}dt'\Bigg\{\varphi_R^{(0)}({\bf x}',t_1)\Gamma''\Big(\varphi^{(0)}_R({\bf x}',t'),t'\Big)\varphi^{(1)}_R({\bf x}',t')\nonumber\\&&~~~~~~~~~~~
-\varphi_L^{(0)}({\bf x}',t_1)\Gamma''\Big(\varphi^{(0)}_L({\bf x}',t'),t'\Big)\varphi^{(1)}_({\bf x}',t')\Bigg\}\,a^3(t')\;.
\end{eqnarray*}
This cancels the $\Gamma''$ terms in the first three lines of Eq.~(66), leaving us with the much simpler relation
\begin{eqnarray}
&& W^{(2)}[J,t_1]_{\rm tree}=i\int_{-\infty}^{t_1}dt\,a^3(t)\int d^3x\;\Bigg\{-\frac{1}{2}\varphi^{(1)}_R({\bf x}, t)\Gamma'\Big(\varphi^{(0)}_R({\bf x}, t),t\Big)\nonumber\\&&~~~~~~~ +\frac{1}{2}\varphi^{(1)}_L({\bf x}, t)\Gamma'\Big(\varphi^{(0)}_L({\bf x}, t),t\Big) \Bigg\}
\;.
\end{eqnarray}
(A similar cancelation occurs in each order of perturbation theory.)  Eqs.~(39) and (40) give the first-order contributions to $\varphi_L$ and $\varphi_R$ for general times:
\begin{eqnarray}
&&\varphi^{(1)}_L({\bf x},t)=\int d^3x'\int_{-\infty}^{t_1} dt'\;\Bigg\{G({\bf x}-{\bf x}',t,t')\,a^3(t')\,\Gamma'\Big(\varphi^{(0)}_L({\bf x}',t'),t'\Big)\nonumber\\&&~~~+G_0({\bf x}-{\bf x}',t,t')\,a^3(t')\,\Bigg[\Gamma'\Big(\varphi^{(0)}_R({\bf x}',t'),t'\Big)-\Gamma'\Big(\varphi_L({\bf x}',t'),t'\Big)\Bigg]\Bigg\}\;,~~~~
\end{eqnarray}
\begin{eqnarray}
&&\varphi^{(1)}_R({\bf x},t)=\int d^3x'\int_{-\infty}^{t_1} dt'\;\Bigg\{G({\bf x}-{\bf x}',t,t')\,a^3(t')\,\Gamma'\Big(\varphi^{(0)}_R({\bf x}',t'),t'\Big)\nonumber\\&&~~~-G^*_0({\bf x}-{\bf x}',t,t')\,a^3(t')\,\Bigg[\Gamma'\Big(\varphi^{(0)}_R({\bf x}',t'),t'\Big)-\Gamma'\Big(\varphi_L({\bf x}',t'),t'\Big)\Bigg]\Bigg\}\;.~~~~
\end{eqnarray}
Using Eqs.~(69) and (70) in Eq.~(68) gives
\begin{eqnarray}
&& W^{(2)}[J,t_1]_{\rm tree}=\frac{1}{2}\int_{-\infty}^{t_1}dt\,a^3(t)\int d^3x\;\int_{-\infty}^{t_1}dt'\,a^3(t')\int d^3x'\;\nonumber\\&&~~~\times \Bigg\{-\Gamma'\Big(\varphi^{(0)}_R({\bf x}, t),t\Big)\Delta_{RR}({\bf x}-{\bf x}',t,t')\Gamma'\Big(\varphi^{(0)}_R({\bf x}', t'),t'\Big)\nonumber\\&&~~~~~+\Gamma'\Big(\varphi^{(0)}_R({\bf x}, t),t\Big)\Delta_{RL}({\bf x}-{\bf x}',t,t')\Gamma'\Big(\varphi^{(0)}_L({\bf x}', t'),t'\Big)\nonumber\\&&~~~~~+\Gamma'\Big(\varphi^{(0)}_L({\bf x}, t),t\Big)\Delta_{LR}({\bf x}-{\bf x}',t,t')\Gamma'\Big(\varphi^{(0)}_R({\bf x}', t'),t'\Big)\nonumber\\&&~~~~~-\Gamma'\Big(\varphi^{(0)}_L({\bf x}, t),t\Big)\Delta_{LL}({\bf x}-{\bf x}',t,t')\Gamma'\Big(\varphi^{(0)}_L({\bf x}', t'),t'\Big)\Bigg\}\;,
\end{eqnarray}
where
\begin{eqnarray}
&&\Delta_{LL}({\bf x}-{\bf x}',t,t')=-iG({\bf x}-{\bf x}',t,t')+iG_0({\bf x}-{\bf x}',t,t')\;,\\
&&\Delta_{RR}({\bf x}-{\bf x}',t,t')=iG({\bf x}-{\bf x}',t,t')-iG^*_0({\bf x}-{\bf x}',t,t')\;,\\
&&\Delta_{LR}({\bf x}-{\bf x}',t,t')=iG_0({\bf x}-{\bf x}',t,t')\;,\\
&&\Delta_{RL}({\bf x}-{\bf x}',t,t')=-iG^*_0({\bf x}-{\bf x}',t,t')\;.
\end{eqnarray}
Referring back to the definitions (35) and (36) of the Green's functions, and recalling the formula (49) for the interaction picture field, we see that Eqs.~(72)--(75) give
\begin{eqnarray}
&&\Delta_{LL}({\bf x}-{\bf x}',t,t')=\langle 0|\bar{T}\{\varphi({\bf x},t)\varphi({\bf x}',t')\}|0\rangle\;,\\
&&\Delta_{RR}({\bf x}-{\bf x}',t,t')=\langle 0|T\{\varphi({\bf x},t)\varphi({\bf x}',t')\}|0\rangle\;,\\
&&\Delta_{LR}({\bf x}-{\bf x}',t,t')=\langle 0|\varphi({\bf x},t)\varphi({\bf x}',t')|0\rangle\;,\\
&&\Delta_{RL}({\bf x}-{\bf x}',t,t')=\langle 0|\varphi({\bf x}',t')\varphi({\bf x},t)|0\rangle\;,
\end{eqnarray}
in agreement with the rules (50)-(53) for propagators in the ``in-in'' formalism.  The signs of the four terms in brackets in Eq.~(71) are just those expected from the rule of associating factors $+i$ and $-i$ with $R$ and $L$ vertices, respectively.  

We have recovered enough of the results of perturbation theory to assure us of the validity of the tree theorem for the ``in-in'' formalism.

\vspace{12pt}

 I am grateful for conversations with P. Greene, E. Komatsu, J. Maldacena, and M. Musso.   This material is based upon work supported by the National Science Foundation under Grant No. PHY-0455649.

\begin{center}
{\bf References}
\end{center}

\begin{enumerate}
\item  J. Schwinger, Proc. Nat. Acad. Sci. US {\bf 46}, 1401 (1961).  Also see K. T. Mahanthappa, Phys. Rev. {\bf 126}, 329 (1962); P. M. Bakshi and K. T. Mahanthappa, J. Math. Phys. {\bf 4}, 1, 12 (1963); L. V. Keldysh, Soviet Physics JETP {\bf 20}, 1018 (1965); P. Danielewicz, Ann. Phys. {\bf 152}, 239 (1984); K Chou, Z. Su, B. Hao, and L. Yu, Phys. Rept. {\bf 118}, 1 (1985); R. D. Jordan, Phys. Rev. D {\bf 33}, 444 (1986); B. DeWitt, {\em The Global Approach to Quantum Field Theory} (Clarendon Press, Oxford, 2003): Sec. 31.  For applications  to cosmology, see E. Calzetta and B. L. Hu, Phys. Rev. D {\bf 35}, 495 (1987); M. Morikawa, Prog. Theor. Phys. {\bf 93}, 685 (1995); N. C. Tsamis and R. Woodard, Ann. Phys. {\bf 238}, 1 (1995); {\bf 253}, 1 (1997); N. C. Tsamis and R. Woodard, Phys. Lett. {\bf B426}, 21 (1998); V. K. Onemli and R. P. Woodard, Class. Quant. Grav. {\bf 19}, 4607 (2002); T. Prokopec, O. Tornkvist, and R. P. Woodard, Ann. Phys. {\bf 303}, 251 (2003); T. Prokopec and R. P. Woodard, JHEP {\bf 0310}, 059 (2003); V. K. Onemli and R. P. Woodard, Phys. Rev. D {\bf 70}, 107301 (2004); T. Brunier, V.K. Onemli, and R. P. Woodard, Class. Quant. Grav. {\bf 22}, 59 (2005); S. Weinberg, Phys. Rev. D {\bf 72}, 043514 (2005); Phys. Rev. D {\bf 74}, 023508 (2006); M. van der Meulen and J. Smit, J. Cosm. Astropart. Phys. {\bf 11},  023 (2007).

\item It is shown by S. Weinberg, Phys. Rev. D {\bf 74}, 023508 (2006) that in a broad class of theories there are no contributions to the correlation functions involving positive powers of the Robertson-Walker scale factor $a$, though powers of $\ln a$ are possible.  The possibility of powers of $\ln a$, arising when the effective cut off provided by renormalization is at virtual physical wave numbers of order $H$, was pointed out by S. Weinberg, Phys. Rev. D {\bf 73}, 043514 (2005).   This was found to occur by  M. van der Meulen and J. Smit, ref. [1].

\item For a non-linear treatment, see J. M. Maldacena, J. High Energy Phys. {\bf 05}, 013 (2005) for the case of single field inflation, and  D. H. Lyth, K. A. Malik, and M. Sasaki, J. Cosm. Astropart. Phys. {\bf 05}, 004 (2005) for  single-field inflation and its aftermath.

\item S. Coleman, in {\em Aspects of Symmetry} (Cambridge University Press, Cambridge, 1985): pp 139--142.

\item  S. Weinberg, Phys. Rev. D {\bf 72}, 043514 (2005).

\item D. Seery, K. A. Malik, and D. H. Lyth, J. Cosm. Astropart. Phys. A {\bf 03}, 014 (2008).

\end{enumerate}

\end{document}